\documentstyle[prd, twocolumn, aps, psfig]{revtex}
\def\be{\begin{equation}}
\def\ee{\end{equation}} 
\def\ba{\begin{eqnarray}}
\def\ea{\end{eqnarray}}         

\tightenlines
\input{psfig}
\begin{document}
\preprint{PRELIMINARY}
\title{Prospects for Constraining Cosmology
with the Extragalactic Cosmic Microwave Background Temperature}
\author{John M.~LoSecco and  Grant J.~Mathews} 
\address{Center for Astrophysics, University of Notre Dame, Notre Dame, 
Indiana 46556}
\author{Yun Wang}
\address{Department of Physics and Astronomy, University of Oklahoma, 
Norman, Oklahoma, 73019}
\date{\today}
\maketitle
\begin{abstract}
Observers have demonstrated that it is now feasible to measure
the cosmic microwave background (CMB) temperature at high redshifts.
We  explore the possible constraints on cosmology which might
ultimately be derived from such measurements.  Besides providing  
a consistency check on standard  and alternative cosmologies, 
possibilities include:  
constraints on the inhomogeneity and anisotropy of
the universe  at intermediate redshift $z\ ^<_\sim 10$;
an independent probe of peculiar motions 
with respect to the Hubble flow; and constraining the epoch of
reionization.
We argue that the best possibility is as a probe of peculiar motions.
We show, however, that the current measurement uncertainty 
($\Delta T=  \pm 0.002\,$K) in the local
present absolute CMB temperature
imposes intrinsic limits on the use
of such CMB temperature measurements as a cosmological
probe.  At best, anisotropies at intermediate
redshift could only be constrained at a level of $^>_\sim 0.1\%$ and
peculiar motions could only be
determined to an uncertainty of $^>_\sim 311\,$ km s$^{-1}$.
If the high $z$ CMB temperature can only be measured
with a precision comparable to the uncertainty of the local interstellar 
CMB temperature, 
then peculiar motions could be determined to an uncertainty
of $1101\,(1+z)^{-1} [\Delta T_{CMB}(z)/0.01\,\mbox{K}] ~{\rm km~s}^{-1}$.

\end{abstract}

\vskip .1 in
PACS numbers: 95.30.Dr, 95.85.Bh, 97.10.Wn, 98.80.-k, 98.80.Bp, 
98.80.Es, 98.58.-w, 98.90.+s

\section{Introduction}

Shortly after the discovery \cite{pandw} of the cosmic microwave background 
radiation (CMB)
it was realized \cite{field,thad} that an extrasolar detection
of this  radiation  had already been made 
at $\lambda=2.63$ mm \cite{mck}
in the relative strengths of the 3874.0 \AA [R(1)] and 3874.6 \AA [R(0)]
absorption transitions of cyanogen (CN) in interstellar  molecular clouds.  
The most recent  measurement of interstellar CN absorption yields
$T_{CMB}=2.729(+0.023, -0.031)\,$K \cite{Roth93,Roth95}. 
This provides an important independent calibration point for
the local CMB temperature, since it measures the background temperature 
far from the solar system.  Indeed, the interstellar 
temperature is in excellent agreement with the
 best {\it COBE} local value
of $2.725\pm 0.002\,$K \cite{Mather99,Fixen96,Mather94}.
It is also  noteworthy that the precision of the interstellar
measurement is now approaching the accuracy of the {\it COBE} measurement.
This raises the interesting question 
as to whether similarly accurate determinations of the CMB temperature
might be possible in extragalactic absorption systems.

\section{The Data}
Indeed, it has now been well demonstrated that
the CMB temperature can be measured at high redshifts 
by using atomic fine-structure transitions in 
cool absorption-line systems along the line of
sight to high-redshift quasars.  This endeavor was first pioneered
by Bahcall and Wolf \cite{BahcallWolf68} who used C\,II excitations of
absorption along the line of sight to PHL 957 to obtain
an upper limit of $T_{CMB} < 45$K at $z = 2.309$.  More recent
investigations have been based upon the $J = 0$, 1, and 2 ground state
fine-structure levels of C\,I.  Among other abundant species
(e.g.~O\,I, C\,II, Si\,II, N\,II, and Fe\,II ), C\,I is 
perhaps best suited because
it has the smallest energy splitting among its fine-structure levels.
A recent summary of the possible absorption lines and their 
relative merits is to be found in \cite{silva00}.

Using C\,I, Songaila et al.~\cite{Song94}, for example,
 have observed  along the line of sight toward QSO 1331+170 
and obtained $T_{CMB}=7.4 \pm 0.8\,$K at $z=1.776$
 consistent with the
expected value of 7.57$\,$K.  Similarly,
Ge, Bechtold and Black \cite{Ge97}  have observed  toward QSO 0013-004 
and obtained $T_{CMB}=7.9 \pm 1.0\,$K at z=1.9731, consistent with the
expected value of 8.1$\,$K.  These measurements, however, must
be taken as upper limits as other excitation mechanisms may have 
contributed to the observed level populations.  Recently, however,
Srianand, Petitjean, and Ledoux \cite{Srian00} have shown 
that an absolute temperature measurement at high redshift is
possible.  In addition to the fine-structure levels of C\,I in 
an isolated remote cloud at $z = 2.338$,  they 
utilized a detection of
several rotational levels in molecular hydrogen to uniquely
constrain competing excitation processes.  
In this way they could deduce
an {\it absolute} temperature of $10 \pm 4$ K, consistent with the expected
value of $9.1$ K. Using a similar technique, Levshakov et al. 
\cite{Sergei} have made a measurement  of $T_{CMB}=
11.7^{+1.6}_{-2.5} $ K  at the highest redshift yet,
 $z = 3.025$.  They also obtained an upper limit of  $T_{CMB} < 15.2$
for a system at $z = 4.466$.

\begin{figure}
\mbox{\psfig{figure=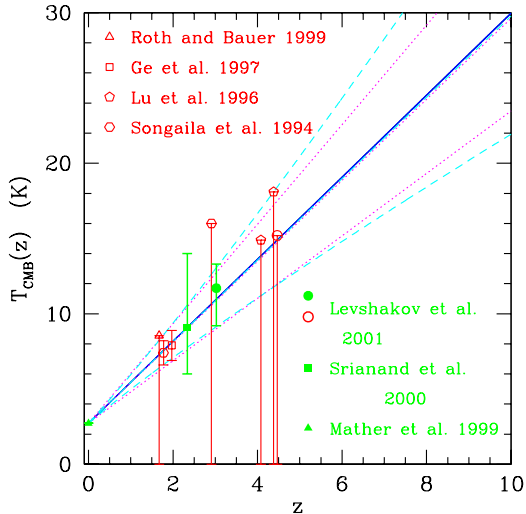,width=3.3in}}
\caption{Summary of some current measurements of the CMB temperature at various
redshifts as labeled. 
 Open symbols are observational upper limits due to the possibility of
non-CMB induced  excitation.  Closed symbols are absolute temperature measurements.
The solid line is the extrapolated {\it COBE} CMB temperature.  Its width is the
extrapolated uncertainty for a FRW universe. The dotted lines are 
the best straight-line fit to the data  and the  $\pm~2 \sigma$ uncertainty.
The dashed lines are for a power-law fit as described in the text.
 }
\protect\label{fig:1}
\end{figure}

A summary of some of the currently available measurements
\cite{Song94,Ge97,Srian00,Sergei,Lu96,Roth99} is 
shown on Figure \ref{fig:1}. At present the
existing measurements
are quite uncertain and most points can only  be treated as upper limits.
Nevertheless,  with forthcoming 
 high resolution spectroscopy on ever larger telescopes, 
we can perhaps anticipate that
accurate determinations of the CMB temperature at 
high redshifts may be possible in the not
too distant future  with a precision comparable to or
better than the present existing  local interstellar
measurement.
Indeed, we may now be
entering  a new epoch in which high precision
measurements of the CMB temperature in the distant past become a 
real possibility.
Hence, in this paper, we discuss how measurements of 
the CMB temperature at various redshifts
might be useful to constrain cosmological models. 
We review the possibilities
and estimate the ultimate viability of such constraints
based upon the current technology for deducing the local CMB 
temperature.

Although the data is sparse, it already constrains 
the possible dependence of the CMB temperature with redshift.
Two sets of lines are drawn on Figure \ref{fig:1}.  One set 
(dotted lines) corresponds to 
a straight-line fit (and $\pm 2 \sigma$ uncertainty) of the form 
\begin{equation}
T(z) = T_0(z)( 1 + b z)~~.
\end{equation}
The other fit (dashed lines) is for a suggested 
\cite{lima} power-law form 
\begin{equation}
T(z) = T(0)(1 + z)^{1 - \beta}~~.
\end{equation}
The fits on these figures are based upon  the {\it COBE} local measurement
plus the two detections
of \cite{Srian00} and \cite{Sergei} and the data of
\cite{Song94,Ge97}.  These latter two points are included as a probable detection, 
though technically 
they are only an upper limit since alternative excitation processes
have not been excluded.  These produce limits of
$b = 0.99 \pm 0.22~(2 \sigma)$ and $\beta  = 0.003  \pm 0.13~(2 \sigma)$,
with $T(0) = 2.725 \pm 0.002 ~(2 \sigma)$ fixed by the {\it COBE} measurement.
If the data set is limited to only the local {\it COBE}  point plus the  two 
firm detections \cite{Srian00,Sergei}, then the limits become
$b = 1.10 \pm 0.55~(2 \sigma)$ and $\beta  = -0.05  \pm 0.25~(2 \sigma)$.
Clearly, even a few more points could significantly  improve
the present limits on variations of the CMB temperature with redshift.

From Figure \ref{fig:1} it is clear that
given the present day CMB temperature,
the standard Friedmann-Robertson-Walker (FRW)
 cosmology makes a precise prediction of 
the CMB temperature versus redshift.
As long as there has been no significant net heating or dust contamination,
and as long as the site of extragalactic CMB measurement is not
affected by a large  gravitational inhomogeneity (see discussion below)
the CMB photon
energies are simply redshifted with the cosmic expansion.
The near perfect plankian  spectral shape of the
local observed CMB places very strong constraints on the
possibility of contamination by an early epoch of violent
star formation or dust.  The current  {\it COBE} limit on the
Compton $y$ parameter, $y = \int (kT_e/m_e c^2) d\tau < 1.5 \times 10^{-5}~
(3 \sigma)$ essentially eliminates the possibility of
any significant distortions
from early star formation/ionization or dust.
 Hence,  to high accuracy one can  invoke a simple relation between 
the cosmic scale factor $a$, the cosmic redshift $z$ and the
expected background radiation temperature $T_{CMB}(z)$ for
photons arriving from
regions that are not too deep within a  gravitational potential,
\begin{equation}
{T_{CMB}(z) \over T_{CMB}(0)} = {a_0 \over a}= 1+z  ~~,
\end{equation}
where $T_{CMB}(0)$ is the present CMB temperature.
An accurate measurement of the present 
CMB temperature thus enables an accurate prediction of the 
FRW  CMB
temperature at all redshifts between now $(z = 0)$
and the surface of last photon scattering $(z \approx 1100)$.

\section{Constraining Alternative Cosmologies}

Although the temperature-redshift relation of the standard hot
big-bang  seems well established, a
direct observation of the correlation of CMB temperature 
with redshift is still a useful cosmological probe.  
At the very least
it confirms the notions of entropy conservation and 
a hotter, denser early universe and helps to eliminate
alternative cosmologies \cite{arp}.  For example,
in principle one might still imagine in spite of a number
of difficulties (cf. \cite{peebles}) that a steady-state 
(or quasi steady-state) cosmology could
be contrived \cite{arp,phillips} to produce a universal 3$K$ microwave by
some combination of starlight and dust.
However, an increase of the  microwave background
temperature with redshift is difficult to achieve in such models.

Nevertheless, alternative models have been
proposed \cite{lima} in which photon creation takes place.  In this
case the temperature-redshift relation would obey
$T_{CMB}(z)/T_{CMB}(0) = (1 + z)^{1 - \beta}$, where $0 \le \beta \le 1$.
These models therefore give a lower temperature at high redshift.
They can only be constrained by  direct measurements of the temperature
redshift relation. As noted above the current $2\sigma$ limit is
$\beta =0.003 \pm  0.13$.  This is consistent with the existing constraint
from big-bang nucleosynthesis \cite{birkel}.  
  
These data also constrain a possible  
Hoyle-Narlikar cosmology \cite{narlikar} in which the galactic redshift
is proposed to result from variations in the electron mass
in a flat Minkowski spacetime.  In this case, the local background
temperature could actually decrease with redshift, though the apparent
temperature would be constant. This also is ruled out
at the level of $8 \sigma$ by the present data.

As a final note, we point out that measurements of the fine structure splitting
at high redshift which determine the CMB temperature 
can also  be used to place limits on the possibility of a time 
varying fine-structure constants \cite{webb00,landau01,Bahcall67}.  
This can happen, for example,
in theories that invoke extra compact dimensions to unify gravity
and other fundamental forces. The cosmological
evolution of the scale factor will then manifest itself 
 \cite{marciano} as a time dependence
of the coupling constants.  Another possibility
is in unified theories which  introduce a new scalar field with couplings
to the Maxwell scalar $F_{a b} F^{a b}$.  The evolution of the scalar field
implies a time variability to $\alpha$ \cite{carroll}.
     
Any time dependence of the fine
structure constant in particular should be apparent in
the observed multiplet splitting at high redshift. 
The relative magnitude $\Delta \lambda/\lambda$  of the multiplet splitting 
scales as
\begin{equation}
{\Delta \lambda \over \lambda} \sim \alpha^2
\end{equation}
 Hence, any limits on the variation of the splitting
with redshift also constrains the fine structure constant,
\begin{equation}
{\Delta \alpha(z) \over \alpha(z)}  \approx {1/2} \biggl[
{(\Delta \lambda /\lambda)_z \over (\Delta \lambda /\lambda)_0} - 1 \biggr]~~,
\end{equation}
where $\Delta \lambda$ is the wavelength difference between 
the observed multiplet lines and $\lambda$ is the weighted mean wavelength of 
the multiplet.
 The  best current limits  \cite{webb00} on a time variation of $\alpha$ 
are at a level of $\vert \Delta \alpha /\alpha \vert \sim 10^{-5}$ based upon
multiplet splitting of different species measured simultaneously.  
The narrow absorption line systems of interest for the CMB measurements
of interest here will require a comparable resolution.  For C\,I,
the absorption multiplets are split by
$\Delta \lambda$ a few hundred $\AA$.  Hence, one must resolve the multiplet 
splitting to $\sim 0.01 \AA$ to achieve comparable accuracy.

\section{Constraining Anisotropy and Inhomogeneity}

Sufficiently accurate CMB temperature measurements 
could possibly probe the inhomogeneity and anisotropy of the
background radiation at redshifts intermediate between the
present epoch and the surface of photon last scattering.
At present our only information on inhomogeneity and anisotropy
of the radiation density are from the surface of last scattering at 
$z \approx 1100$.  

First, imagine an idealized case in which
one could make a number of measurements
along nearly the same line of sight but at different
redshifts, $T_{CMB}(z)$.  This might be possible, for example,
if enough narrow absorption-line
systems exist along the line of sight to a distant quasar.
 If no other effects were in operation,
then the difference in temperature at some location
compared with the expected mean FRW temperature $\langle T_{CMB}(z) \rangle$
could be taken as a measure of the inhomogeneity in radiation 
density along that line of sight,
\begin{equation}
{T_{CMB}(z) - \langle T_{CMB}(z) \rangle \over \langle T_{CMB}(z) \rangle} \sim
{\Delta \rho \over \rho}~~.
\end{equation}
 Although  in this case one mainly measures 
the inhomogeneity of the universe, it is dependent upon several 
complicating factors as described below.

Similarly, in another idealized situation,
assume we have high quality measurements of $T_0$, and $T_i(z)$
where $T_i(z)$ refers to numerous measurements of the background temperature
at different points on the sky, but the same observed redshift.
There are then at least two more kinds of test that one could do:

One is to measure the  mean temperature $\bar T(z)$,
where,
\begin{equation}
\bar T(z) = {1 \over N} \sum_{i=1,N} T_i(z)~~.
\end{equation}
On average one should expect some of the deviations 
listed below to cancel.   If enough
points could be obtained at a given $z$, this quantity should
agree closely with the FRW prediction.
Thus,  $\bar T(z)$ as a function of $z$ tests the thermal history of the universe.
Any deviation of this quantity from the FRW prediction could
require a new cosmological paradigm.

Another possible test is the difference $\Delta T(z) = T_i(z) -  \bar T(z)$
at fixed $z$.  The difference of the radiation
density at distinct points on the sky but fixed $z$  is a measure
of some combination of inhomogeneity and anisotropy.
As is easily seen from  Figure 2, different regions of 
the last scattering surface are
being seen at different redshifts, and thus,
 the anisotropy
of the last scattering surface as seen at points  $z_i$ 
is tested.  However,  the surface of constant $z$ is
not necessarily parallel to the surface of last scattering as illustrated
in Figure 2.  This measure of inhomogeneity and isotropy
may be influenced by other factors such as those listed below.

\section{Effects on the Observed  $z-T$ relation}
\begin{figure}
\psfig{figure=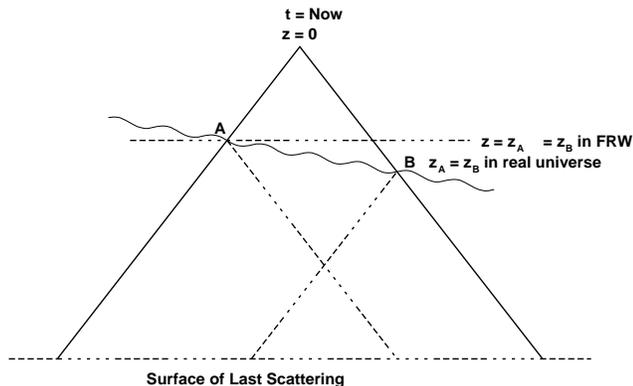,angle=0,width=3.3in}
\caption{Schematic illustration of the lightcone in a perturbed
universe vs. a pure FRW universe.
 }
\protect
\label{lightcone}
\end{figure}

$\mbox{From}$ the above discussion, it is clear that one must carefully  identify the
dominant influences on the possible deviations of the local microwave temperature
from one location to another.  
The true situation is more complicated than the simple FRW picture
especially with regard to detecting any
angular anisotropy.  We illustrate this in the 
 lightcone structure sketched in
Figure \ref{lightcone}.
$\mbox{From}$ the point of view of observations, it is  
natural to work with the measured constant-$z$  hypersurfaces
sketched by the wavy line on Figure \ref{lightcone}.
The points $A$ and $B$ represent two points on such
a hypersurface,  specified by $z=z_A = z_B$ and the directions
$\theta_A$, $\phi_A$ and $\theta_B$, $\phi_B$  on the sky. 

In an unperturbed FRW universe the hypersurfaces of constant 
$z=z_A = z_B$ and constant
energy density are obviously equivalent. This is no longer true in
a universe perturbed by for example:  (i) intrinsic temperature inhomogeneities of the 
surface of last scattering;
 (ii) inhomogeneities of the local gravitational potential at $z_A$ or $z_B$;  
(iii) inhomogeneities due to
the integrated Sachs-Wolfe effect between the surface of last
scattering and  $z_A$ or $z_B$;
 and  (iv) peculiar velocities  at $z_A$ or $z_B$.  
This means that in the real universe, the $z=\,$constant and the last scattering
surface (to a first approximation constant energy density) are not
``parallel" to each other, i.e.~in Figure \ref{lightcone} the points A and B are at the
same observed redshift.
We digress here to discuss the likely best obtainable measurement
error and how this compares with these  possible deviations from the
expected FRW $z-T$ relation.

\subsection{Measurement Error}
For a measurement error comparable to that of the local interstellar
 CMB measurements, $\Delta T_{CMB}(z) \sim \pm 0.02\,$K,
one might expect to constrain the anisotropy and inhomogeneity of the universe
at a level of 
\ba
\frac{\sigma_{T_{CMB}(z)}}{\langle T_{CMB}(z) \rangle} &>& 
\frac{\Delta T_{CMB}(z)}{\langle T_{CMB}(z) \rangle}
\nonumber \\
&=&
\frac{\Delta T_{CMB}(z)}{(z+1) T_{CMB}(0)}= 7 \times 10^{-3}/(z+1)~~,
\ea
corresponding to a level of about 0.2\% at a redshift of 3.
Even if the only uncertainty could be reduced to
the present measurement error of the local CMB temperature propagated
to redshift $z$,  the best one could hope to do
corresponds to a present limit of about 0.1\%, i.e.
\ba
\frac{\sigma_{T_{CMB}(z)}}{\langle T_{CMB}(z) \rangle} &>& 
\frac{(1+z) \Delta T_{CMB}(0)}{\langle T_{CMB}(z) \rangle}
\nonumber \\
&=&
\frac{\Delta T_{CMB}(0)}{T_{CMB}(0)}=7.34 \times 10^{-4}.
\ea
This uncertainty is to be compared with the possible
deviations from the pure FRW picture.

\subsection{Intrinsic Temperature Fluctuations on the
Surface of Last Scattering}
There is no reason
in standard cosmological models
to expect the temperature fluctuations  at intermediate
redshift to substantially differ from the presently observed
fluctuations in the CMB temperature of 
$\Delta T_{CMB}(0)/T_{CMB}(0) \sim  10^{-5}$. They are therefore
probably not detectable.  CMB temperature measurements
at various redshifts, however, 
might still constrain exotic cosmological models in which the universe 
could be postulated to have experienced 
substantial entropy gain or loss, and/or oscillations 
at some  intermediate ($z ^<_\sim 10$) epoch. 

\subsection{Gravitational Potential at the Absorber}

  As noted in the introduction, the best place in
which to detect the local extragalactic background temperature
is probably in cool, narrow absorption-line systems along the
line of sight to background quasars.  Generally, it is believed \cite{hernquist}
that  such systems reside in the intergalactic medium
 in large filamentary  flattened structures of low or  moderate overdensity.
 Hence, the local gravitational redshift is probably negligible for such
systems.

 Nevertheless,
one should keep in mind the magnitude of a possible gravitational
redshift $ \Delta T_g /T (z) = z_g \approx   G M/Rc^2$  which 
might alter the $z-T$ relation.  
Even for the extreme case of a cloud within  
a galaxy with $M \approx 10^{12}$ M$_\odot$ which has been 
compressed to $R \sim 1 $ kpc,
the gravitational redshift is only $z_g \sim 5 \times 10^{-5}$ and thus 
well below our estimated best observable limits.
Similarly,
for a large dense galactic cluster one might envision 
a worst case of a cloud at a distance of only $R \sim 1 $ Mpc
from a compact galactic  cluster of $M \approx 10^{14}$ M$_\odot$.
 Even for this extreme  case 
the redshift effect is only $z_g \sim 5 \times 10^{-6}$, again well
below our estimated detection limits.  

\subsection{Integrated Sachs-Wolfe Effect}

For photons which propagate from the surface of last scattering to
point $z_A$ or $z_B$ in Figure 2,  there are two ways
to think of the Sachs-Wolfe effect \cite{sachswolfe}.  If large-scale departures
from homogeneity caused the expansion of the universe to
differ along different lines of sight, then this would
produce a quadrupole anisotropy.  The current  {\it COBE} limits
on the CMR quadrupole anisotropy, however, limit this 
possibility to an expected amount of $\Delta T/T \sim 10^{-5}$.
Hence, this is probably not detectable.  

Alternatively, one can view this effect as a correction for the 
gravitational potential of the mass fluctuation spectrum 
at the surface of last scattering.  This gravitational
redshift is expected to be much smaller than the gravitational
redshifts described above, i.e. comparable to the intrinsic
temperature fluctuations at the surface of last scattering.
Indeed, for absorbers at  $z > 1$, we may expect that this
contribution can be neglected.  
Similarly,  the effects
of photons crossing gravitational inhomogeneities between the surface
of last scattering and the absorber, or between the absorber 
and the observer, cancel unless the fluctuations are
comparable to a Hubble length.

  Hence, we conclude that the effects of 
inhomogeneities in the absorber gravitational potentials or the surface
of last scattering are not likely to be detectable by
measurements of the CMB temperature at high redshift.
Indeed, the only possible observable effects are
probably those due to streaming motions, which we
now consider in more  detail.

\section{Probes of Large-Scale Motion}

\subsection{Dark-Matter Potentials}
Perhaps a more useful cosmological constraint
could come from using the CMB temperature
at intermediate redshifts to probe large-scale peculiar motion.
Over 90\% of the matter in the universe is dark (invisible).
It only manifests itself through gravity. It is important, therefore,
to constrain the dark matter potentials through peculiar (i.e.,
local) motions of galaxies and clusters of galaxies.
The matter density fluctuations $\delta$ are related to the peculiar 
velocities ${\bf v}_p$ locally by
\be
{\bf \nabla \cdot  v}_p= -f(\Omega)\, \delta,
\ee
where $f(\Omega)=d\,\mbox{ln}\delta/d\,\mbox{ln}a \simeq \Omega_m^{0.6}$, 
with $a$ denoting
the cosmic scale factor. In principle, it is possible to
reconstruct the mass density field from observed peculiar
velocities \cite{Dekel93}. 
If peculiar velocities are sufficiently well known,
then statistical studies of the peculiar velocity distribution
in the universe can be used to constrain cosmological models.

 In principle, such peculiar velocities can be determined
from measurements of the CMB temperature.
If measurement of the CMB temperature at a given redshift
can be made sufficiently precise, 
a comparison of the spectroscopic redshift to the CMB temperature 
in a distant absorption-line system could be used
to detect the line-of-sight component of the absorber's peculiar
motion.

The observed spectroscopic redshift $z_s$ for a CMB absorption 
system is actually the combination of two effects.  That is,
$1+z_s=(1+z)(1+z_p)$, 
where $z$ is the cosmic redshift, while $z_p$ is the redshift due
to the component of peculiar velocity along the observed line of sight
\cite{Peacock99}.
Although the net spectroscopic redshift is  
dependent upon proper motion,
the deduced CMB temperature is almost independent of proper motion.
The only effect of proper motion on the background temperature
is to produce a dipole moment which
averages out in the angle-integrated net population temperature.
Thus, the CMB temperature can be used to deduce the
true cosmological redshift via equation (1), $z=T_{CMB}(z)/T_{CMB}(z=0) -1$. 
The  peculiar velocity  is just 
c$z_p = c(z_s -z)/(1+z)$, so that its uncertainty can be written:
\ba
c\Delta z_p&=& c\, (1+z)^{-1}
\sqrt{ (1+z_p)^2 (\Delta z)^2 +(\Delta z_s)^2 }\nonumber  \\
&=& c \biggl\{ (1+z_p)^2 \left[
\left({\Delta T_{CMB}(z) \over T_{CMB}(z)}\right)^2 
 +
\left(\frac{\Delta T_{CMB}(0)}{T_{CMB}(0)}\right)^2 \right]
 \nonumber \\
 &&+(\frac{\Delta z_s}{1+z} )^2 \biggr\}^{1/2},
\ea
where $\Delta T_{CMB}$ is the uncertainty in the measurement of
$T_{CMB}$, and $\Delta z_s$ is the uncertainty in the
spectroscopic redshift. 
The error in the observed spectroscopic redshift is usually
much less than the error in the CMB temperature and
can be neglected.  If we assume that a precision comparable to
the uncertainty from the local interstellar CMB temperature
can be obtained, then this implies a limit to detectable velocities of,
\begin{equation}
c\Delta z_p\ ^>_\sim  \,
\frac{1101}{(1+z)}\, 
\left[\frac{\Delta T_{CMB}(z)}{0.01\,\mbox{K}}\right] ~{\rm km~s}^{-1}.
\end{equation}
Note, that for a given measurement uncertainty $\Delta T_{CMB}(z)$,
it is {\it easier} to detect peculiar velocities at large $z$,
since the Doppler shifted wavelength of the photons, $(1+z_p)\lambda_0$,
is shifted an additional factor of $1+z$ due to the expansion
of the universe.  This effectively amplifies the present signature of
the peculiar motion $v=cz_p$ at cosmic redshift $z$.

In the limit that the  measurement uncertainty could
be reduced to as small as the current {\it COBE} uncertainty in the
local CMB temperature, we would still  have
$\Delta T_{CMB}(z) \ge (1+z) \,\Delta T_{CMB}(0)$, hence
\begin{equation}
c\Delta z_p\ ^>_\sim  311\,{\rm km~s}^{-1}~~.
\end{equation}

Thus, only large peculiar motions are likely to be detectable by this
technique.  Nevertheless,
there are significant large-scale motions in
the universe \cite{rubin}. In the central parts of rich superclusters,
the line-of-sight peculiar velocities of clusters are of order
1200$\,$km s$^{-1}$ and can be much 
larger for individual galaxies \cite{Bahcall}. 
Such clusters are important as tracers
of the gravitational potential of the superclusters.
The excitation temperature of absorbers in these systems
might therefore provide useful constraints on the radial peculiar velocities.

\subsection{Distance  Calibration}

The CMB temperature at small redshifts
may also provide support for distance standards.
To establish distance ladders to cosmological distances, observers need to
calibrate Cepheid variables in nearby galactic clusters. One uncertainty
in such calibration
has been the relative positions of the Cepheid host galaxies
within the clusters.
The maximum peculiar velocity corresponds to galaxies
located in the center of the cluster.  Therefore, 
if one can measure the peculiar velocity of the host galaxies to
sufficient accuracy via the combination
of the measurements of the CMB temperature and redshift 
at the location of a Cepheid host galaxy,
one could in principle eliminate the ambiguity in
the calibration
of Cepheids arising from its location relative to the center of the cluster.

\section{Constraining the Epoch of Reionization}

The universe was probably reionized at $z ^>_\sim 3$.
The time scale over which the universe made the transition from being
neutral to being almost completely reionized is unknown. Constraints
on reionization will help reveal the nature of the first generation of 
objects that ended the so-called dark ages of the universe.
Radio observations of the redshifted 21-cm emission
of neutral hydrogen have been proposed \cite{Tozzi99} as a means
to search for the signature of reionization.  Basically,
one expects a sharp cutoff in the 21-cm emission for the redshift interval
corresponding to the epoch of reionization.

We point out that similarly,
there should be a sharp drop in the abundance of the neutral
atomic and diatomic species of interest here 
at the epoch of reionization.
It is likely, therefore, that  CMB-induced C\,I, O\,I, CN and CH excitations 
might be observed at high redshifts
(before reionization) and also after reionization
at intermediate redshifts (say, $z\sim 2$) but not in between.
Whereas the signal from ionized species like C\,II, Si\,II,
N\,II, and Fe\,II may actually increase.
Since all of these absorption lines occur at
wavelengths around 4000$\,\AA$ in the rest frame, they will be
redshifted to the infrared for photons absorbed during the reionization epoch.
NGST will be the ideal instrument for detecting  such features
over a wide redshift range. The existence of a gap
in redshift space for the appearance of such features might be
an independent  means to identify the epoch of reionization.

\section{Observational Feasibility}

\subsection{Diatomic Absorption Features}

While using the atomic fine-structure transitions in C\,I is a proven
method of measuring the CMB temperature at high redshifts 
\cite{Song94,Ge97,Srian00},
the use of the transitions to rotational or fine-structure modes
in the diatomic molecules CN, CH, and CH$^+$ might still deserve further
exploration.  Although less abundant,
they may provide an alternative and complementary 
method of measuring the CMB temperature at low and high redshifts.
Since they are the preferred absorption features for the local interstellar
medium, one might take advantage of ratios to the local interstellar absorption
features to minimize systematic error.  Hence, we summarize here some
of the observational considerations.

The observed equivalent widths of a pair of lines ($i$, $j$) in a
molecular cloud, 
$W_i$ and $W_j$,
are converted into column densities $N_i$ and $N_j$
assuming a single-component Gaussian curve of growth
in each case. Any unresolved structure in the lines will be
accurately represented by this approach, provided there are no
narrow, heavily saturated, optically thick components whose
column densities dominate the composite line-of-sight value
\cite{Roth95,Spitzer78,Jenkins86}.
Assuming thermal excitation by the CMB, the ratio of the column densities
is given by a Boltzmann factor:
\be
\frac{N_i}{N_j} = \frac{g_i}{g_j}\, \exp\left(
-\frac{h \nu_{ij}}{k_B T_{CMB}}\right),
\ee
where $g_{i,j}$ are statistical weights of the lower and upper
states, and $h \nu_{ij}=E_i-E_j>0$ is the difference in energy
of the two rotational (CN) or fine-structure (CH) states.

To make predictions for the observable equivalent widths 
at a redshift $z$ for an arbitrary transition $i \rightarrow j$
due to the CMB, we write
\be
\frac{N_i}{N_j} = \frac{g_i}{g_j}\, \exp\left(
-\frac{1\,\mbox{mm}}{\lambda_{ij}}\,
\frac{5.278}{1+z}\right),
\ee
where $\lambda_{ij}=c/\nu_{ij}$, and we have used $T_{CMB}(z)
=2.728 (1+z)$.
To a good approximation, we can use
$W_i \propto N_i f_i \lambda_i^2$, where $f_i$ is the oscillator
strength of the transition, and $\lambda_i$ is the wavelength
of the absorption line. Now we find
\be
\label{eq:Wij}
\frac{W_i}{W_j} = \frac{g_i}{g_j}\, \frac{f_i}{f_j}\,
\left(\frac{\lambda_i}{\lambda_j}\right)^2\,\exp\left(
-\frac{1\,\mbox{mm}}{\lambda_{ij}}\,
\frac{5.278}{1+z}\right).
\ee
Clearly, observation of the absorption line from the excited
state $i$ is favored by large $\lambda_{ij}$. The smaller 
$\lambda_{ij}$, the smaller $W_i$ relative to $W_j$,
the more suppressed the absorption line $i$
relative to absorption line $j$. This explains the relative
strength of the CN lines (in order of decreasing line strength)
R(0), R(1), and the marginal
detection of the R(2) line,
and the non-detection of the CH line R$_1$(1)
in local measurements \cite{Thaddeus72}.

In general, the relative strength of an absorption line from
an excited state increases with redshift $z$. Since all of
the absorption lines in the strongest interstellar bands of CN
have wavelengths of  about 4000$\,\AA$,
we are restricted to $z <2$ by the wavelength ($\sim 1\,\mu$m) at which 
optical observations are hindered by 
atmospheric absorption. However,
space-based infrared observations from HST 
or the upcoming NGST can search
for CN and CH excitations at redshifts $z ^>_\sim 3$. 

For CN, we find $W_1/W_0= 2.00\, \exp[-2/(1+z)]$. This
gives $W_1/W_0 = 0.27$, close to the observed value of 0.31
for $\zeta$ Ophiuchi. For $z=1$, $W_1/W_0 = 0.74$;
for $z=2$, $W_1/W_0 = 1.03$. These estimates agree with
numerical results. Thus, prospects for
observing CN excitation at large redshifts is rather good,
as long as CN clouds can be found at these redshifts.
Although CN absorption from an excited rotational state
has not yet been detected outside the Milky Way,
observational efforts should be directed toward their detection, since
CN has been demonstrated to be an excellent CMB thermometer
\cite{Roth93,Roth95,Palazzi92}. Using an approach first explored by 
Penzias, Jefferts and Wilson \cite{Penzias72}, Roth et al. \cite{Roth93,Roth95}
directly measured the amount of local
excitation from millimeter observations of CN rotational line
emission. Once local effects were reliably understood, they were able
to make small corrections to the CN excitation which has led to the
current accurate measurement of the interstellar CMB temperature of
$T_{CMB}=2.729(+0.023, -0.031)\,$K \cite{Roth93,Roth95}.

For CH, the exponent in Eq.(\ref{eq:Wij}) leads to a factor
of suppression of 8$\times 10^{-5}$ for the equivalent width
ratio of R$_1$(1) and R$_2$(1). The excitation wavelength
for these two states is 0.56$\,$mm, while the CMB emission
at $z=0$ peaks at $\lambda=1\,$mm, so the nearby CH clouds
miss about 10\% of the CMB intensity. It is not surprising, therefore,
that the CH excitation has not been observed locally.
At $z=1$, the equivalent width ratio of R$_1$(1) and R$_2$(1)
increases by a factor of 113. Also, the excitation wavelength
0.56$\,$mm becomes well matched with the wavelength
at which the CMB emission peaks at $z=1$.
Therefore, $z=1$ may be the optimal redshift at which to measure
for the CH excitation and associated CMB temperature.
Similar arguments can be made about the CH$^{+}$ excitation.

\subsection{Finding Cold Absorption-Line Systems}

If one is to look absorption in any of the proposed 
atomic and molecular species, 
one must of course find cold clouds in which the 
these absorbers could exist in significant population.  Although 
such cold clouds are now known to exist \cite{Srian00,rauch98} 
they are quite rare among Lyman alpha systems. Hence, one may wish
to consider other possible systems in which such detections could
be made.

Among possible objects for extragalactic measurements of the CMB temperature,
some promising, as yet unexplored, candidates come to mind.  
One might, for example,  examine the cool low-density
 gas in the external regions of
spiral galaxies  which happen to lie along the line-of sight to
a bright background galaxy or quasar. 
Another possibility might be to study gravitationally lensed
quasars (cf. \cite{rauch98}).  Since the light from such QSO's must 
pass around the lensing galaxy one could at least envision that 
some intervening cool absorption clouds may lie along the path
in the outer regions of the lensing galaxy.  

Obviously, one must consider extremely narrow absorption-line
systems.
Since, for example,
 the rotationally excited line in CN  lies only about 0.63 \AA~below the
ground state,  absorption-line dispersion in the source can lead to smearing
that make the determination of the ratio of equivalent widths
less precise,
($\Delta \lambda / \lambda = 1.6 \times 10^{-4}$).
A 50 km s$^{-1}$ Doppler shift is enough to smear the lines.  A number of
solutions to this trouble may
be possible.  For a galactic source, the effects of galactic 
rotation can be reduced by either
restricting the measurement to galaxies viewed face on or by masking out
a small portion of the galaxy image where the velocity dispersion is
smaller.  In some cases one may be able to measure the dispersion from the
shape of nearby well known singlet lines and use the measured dispersion to
extract the ratio of absorption intensities from the smeared data in the
region of the absorption lines of interest.

\section{Summary}

The uncertainty in both the  locally  and extragalactic
measured CMB temperature place 
intrinsic limits on their usefulness as a means to
constrain cosmological models.
Here we speculate that the use of very high resolution spectrometers 
on large aperture telescopes might facilitate a  1-2 order of magnitude 
improvement in the CMB temperature measurement at high redshifts 
(i.e., comparable to the accuracy of determining the CMB temperature
from the local interstellar medium).  Such accurate
observations would enable us to constrain the anisotropy
and inhomogeneity of the universe on the level of 0.2\% 
(the intrinsic limit is 0.1\%) out to redshifts of a few.

Sufficiently accurate measurements of the CMB temperature at
various redshifts might also 
be a useful probe of large-scale motion in the universe.  
If the high $z$ CMB temperature can be measured
with a precision comparable to the uncertainty of the local interstellar 
CMB temperature, 
then peculiar motions can be determined to an uncertainty of 
$1101\,(1+z)^{-1} [\Delta T_{CMB}(z)/0.01\,\mbox{K}] ~{\rm km~s}^{-1}$,
which can place useful constraint on cosmological models.
The current measurement uncertainty 
($\Delta T=  \pm 0.002\,$K) in the local
present absolute CMB temperature
imposes an ultimate resolution limit of at least  $311\,$km s$^{-1}$.
Nevertheless, even at this level of accuracy,
measurements of the CMB temperature at low reshifts might 
be used to independently calibrate distance standards.

We argue that further
observational efforts should be directed toward 
high precision searches 
for the fine-structure excitations of
atomic C\,I, O\,I, C\,II, Si\,II, N\,II, and Fe\,II
and rotational excitations of  CN, CH, and CH$^{+}$ 
in extragalactic systems.  Even a few more observational
points could significantly constrain some alternative cosmologies
and also provide complementary determinations of the 
possible time variability of
the fine structure constant at the same time.
CMB induced excitations remain good (even better)
thermometers out to large redshift.
If these CMB induced excitations can be found at redshifts
beyond a few through spaced based infrared observations, 
they may also
provide useful constraints on the epoch of reionization.

We thank Terry Rettig for helpful discussions concerning 
absorption-line observations.
Work supported in part by DOE Nuclear Theory grant DE-FG02-95ER40934 at UND
and NSF CAREER grant AST-0094335 at UOK.

\end{document}